%% file: main.tex
\documentclass[preprint,usenatbib,twocolumn]{aastex63}

\usepackage{newtxtext,newtxmath}
\usepackage[T1]{fontenc}
\usepackage{ae,aecompl}
\usepackage{graphicx}	% Including figure files
\usepackage{amssymb}	% Extra maths symbols
\usepackage[utf8]{inputenc}
\usepackage{mathtools}
\usepackage{color}      
\usepackage[caption=false]{subfig}
\usepackage{float}
\usepackage[normalem]{ulem}
\usepackage{subfiles}
\usepackage{natbib}
\received{XXX}
\revised{XXX}
\accepted{XXX}
\submitjournal{APJ}
\shortauthors{Arora et al.}

\begin{document}
\title{Power law pseudo phase-space density profiles of dark matter halos: fluke of physics?}

% The list of authors, and the short list which is used in the headers.
% If you need two or more lines of authors, add an extra line using \newauthor
\author{
Arpit Arora}
\affiliation{School of Physics and Astronomy, University of Minnesota, \\
116 Church Street SE, Minneapolis, MN 55455, USA
}
\author{Liliya. L. R. Williams}
\affiliation{School of Physics and Astronomy, University of Minnesota, \\
116 Church Street SE, Minneapolis, MN 55455, USA
}

\begin{abstract}
It has been known for nearly 20 years that the pseudo phase-space density profile of equilibrium simulated dark matter halos, $\rho(r)/\sigma^3(r)$, is well described by a power law over 3 decades in radius, even though both the density $\rho(r)$, and the velocity dispersion $\sigma(r)$ deviate significantly from power laws. The origin of this scale-free behavior is not understood. It could be an inherent property of self-gravitating collisionless systems, or it could be a mere coincidence. To address the question we work with equilibrium halos, and more specifically, the second derivative of the Jeans equation, which, under the assumptions of (i)  Einasto density profile, (ii) linear velocity anisotropy - density slope relation, and (iii) $\rho/\sigma^3\propto r^{-\alpha}$, can be transformed from a differential equation to a cubic algebraic equation. Relations (i)-(iii) are all observed in numerical simulations, and are well parametrized by a total of 4 or 6 model parameters. We do not consider dynamical evolution of halos; instead, taking advantage of the fact that the algebraic Jeans equation for equilibrium halos puts relations (i)-(iii) on the same footing, we study the (approximate) solutions of this equation in the 4 and 6 dimensional spaces. We argue that the distribution of best solutions in these parameter spaces is inconsistent with $\rho/\sigma^3\propto r^{-\alpha}$ being an fundamental property of gravitational evolution, and conclude that the scale-free nature of this quantity is likely to be a fluke.
\end{abstract}

\section{Introduction}
\subfile{Sections/introduction.tex}

\section{Summary of our analysis methods}\label{outline}

\subfile{Sections/outline.tex}

\section{Anisotropic constrained Jeans (ACJ) equation}\label{CJEsection}

\subfile{Sections/ACJE.tex}

\section{Conclusions}\label{Conc}

\subfile{Sections/Conclusion.tex}

\vskip0.2in
\bibliography{mybib}{}
\bibliographystyle{aasjournal}

\end{document}

%% file: Sections/introduction.tex
Dark matter, the dominant mass component of the Universe, is the scaffolding that provides the structure for galaxies and clusters of galaxies. Therefore, understanding the structure of dark matter halos is one of the most important goals of modern cosmology. While the equilibrium structure of stars has been known for about a century, the structure of equilibrium dark matter halos has proven harder to establish. Using the equation of hydrostatic equilibrium and the equation of state for gaseous material allows one to solve for the internal structure of stars. In the case of collisionless dark matter halos, it is not possible to solve the hydrostatic equilibrium equation---also known as the Jeans equation---because the equation of state of dark matter is unknown\footnote{The equation of state referred to here and other similar contexts in the literature is not the same as the relativistic equation of state, $w=0$, for cold dark matter, used in the cosmological context. Here, the equation of state refers to the relation between mass density $\rho$, and dynamically generated "pressure" $P$ of dark matter, where $P=\rho\sigma^2$, and $\sigma$ is the velocity dispersion.}. 
 
 In 2001, an interesting observation was published by \cite{Tay01}. Using N-body simulations, the authors measured mass density, $\rho(r)$, and velocity dispersion $\sigma(r)$, profiles of equilibrium dark matter halos. The quantity $\rho/\sigma^3$ turned out to be a power law in radius over about 3 decades, despite the fact that neither density nor the velocity dispersion are power laws, but in fact significantly deviate from a scale-free form. Since then, it has been confirmed by a number of studies that dark matter halos formed in cosmological N-body simulations \citep{Dra17,But16,Nol16,Gao12,Lud10,Nav10,Ma09,Kno08,Hof07,Pei06}, 
 halos computed through an iterative collisionless spherical collapse \citep{Aus05}, and even galaxies and clusters formed in the real Universe \citep{Cha14,Mun14} are well characterized by a power law,
 \begin{equation}\label{ppsd}
 \frac{\rho(r)}{\sigma^3(r)}\propto r^{-\alpha}.
 \end{equation} 
 Because this quantity has the dimensions of phase-space density, it has been nicknamed pseudo phase-space density. 

Several papers made attempts to shed light on its origin \citep{Ala13,Lud11,Hen06,Aus05}. In the meantime, others continued to address the more general question of how to understand the structure of dark matter halos that develops so robustly in simulations \citep{Ber19,Pon13,Sal12,Kan11,He10}. \cite{Hjo10} proposed a theoretical derivation for the differential energy distribution of self-gravitating relaxed collisionless matter. Based on the principles of statistical mechanics, they proposed the most likely steady-state configuration of these systems. Their result, DARKexp, forms a one shape parameter family, with $\phi_0$ characterizing the dimensionless depth of the central gravitational potential. DARKexp gives very good fits to the density profiles \citep{Nol16,Hjo15} and, more importantly, to the differential energy distributions of simulated dark matter halos \citep{Wil10b,Nol16}. It also fits well the density profiles of observed equilibrium galaxy clusters \citep{Ber13}. It was shown in \cite{Wil10} that the $\rho/\sigma^3$ profiles of the DARKexp family are close to, but not exactly power laws for many values of $\phi_0$, suggesting that it may not be a universal feature of relaxed systems. Hints of non-universality of this quantity have also been noted in other papers \citep{Del15,Del11,Ma09}.

Recently, \cite{Nad17} challenged the physical origin of the pseudo phase-space density. They consider 1D self-similar fluid collapse, following closely an earlier study by \cite{Ber85}. The authors follow the evolution of gas entropy, whose definition is effectively the same as that of the pseudo phase-space density,  $\rho/\sigma^3$. Because they are dealing with gas, their treatment cannot incorporate velocity anisotropy, which is measured to be non-zero in numerical dark matter simulations \citep{Han06a,Han06b,Lem12}, as well as some observations of galaxies and galaxy clusters \cite{Han07,Han11,Lon18}, and stellar and globular cluster populations in the Milky Way \cite{Vas19}.

This paper is a further attempt to understand $\rho/\sigma^3$: is there some physical principle behind its power law nature, or is it a mere coincidence. Demonstrating the existence of an underlying physical principle will have important implications for our understanding of self-gravitating collisionless systems. { Here, we do not address the possible physical meaning of $\rho/\sigma^3$, but instead assume that if one exists, the final equilibrium state of halos will satisfy eq.~\eqref{ppsd}.}

Our approach differs significantly from that of \cite{Nad17}. While these authors dealt with evolution of isotropic gaseous material, the present work { does not consider halo evolution, and instead} concentrates on the equilibrium state of halos, where velocity anisotropy plays an important role. Though our methods are very different, our conclusions are essentially the same as theirs: pseudo phase-space density is unlikely to have fundamental physical interpretation, and hence cannot help in the understanding of dark matter halos.

%% file: Sections/outline.tex
Because we are dealing with equilibrium dark matter halos, the starting point of our analysis is the Jeans equation, a statement of hydrostatic (or, mechanical) equilibrium for collisionless matter.  Based on the results of N-body simulations,  all the quantities characterizing the spherically averaged equilibrium halos, namely (i) the density profile, (ii) velocity anisotropy profile, and (iii) pseudo phase-space density profile can be modelled, to a good approximation, as simple analytic relations, with a total of 4 or 6 parameters, depending on how the density
profile is represented.

One can make even stronger statements regarding (i) and (iii), going beyond fitting functions to simulations. 

The density profiles of relaxed systems are given by a theoretically derived DARKexp, whose radial profile shape is known exactly. For $\phi_0\approx 4.5$, this shape is very closely matched by Einasto profiles, which is why Einasto fit simulated halos very well. In section~\ref{solve1ACJE} we represent density profiles by Einasto profiles, giving us a total of 4 models parameters for (i)-(iii). In section~\ref{solve3ACJE} we assume the density profiles can come from a wider range of DARKexp models, and represent them with 3 Einasto segments, parametrized by 3 parameters, giving us a total of 6 model parameters for (i)-(iii).

If eq.~\eqref{ppsd} is a robust property of collisionless Newtonian gravity, then one expects this form to be very closely adhered to by equilibrium dark matter halos. Put differently, it would be pointless to try to explain the radial dependence of $\rho/\sigma^3$ if it is not a nearly exact power law. In this paper we do not study dynamical evolution of halos; { only their final equilibrium configurations.} We explain the principle behind our analysis later in this section.

The only one of the relations (i)-(iii) that does not have a nearly exact form is the velocity anisotropy profile, (ii). Here we have to rely on simulations. {Fortunately, as we show in section~\ref{grid} the conclusions of our analysis do not depend on the exact shape of this relation.} \cite{Han06a} and \cite{Han06b} have shown based on a variety of initial conditions, that equilibrium halos have a tight linear relation between velocity anisotropy $\beta$, and the double logarithmic density profile slope, $\gamma$. \cite{Han06a} give a range of parameters characterizing that relation, while \cite{Han06b} present a single set of best fitting parameters. We use both of these results in our analysis. 

At the end of their dynamical evolution, simulated halos attain a state where all three relations, (i)-(iii), have \emph{nearly} exact parametric forms, represented with 4 or 6 parameters. The values of these parameters have been measured from simulations; let us call them collectively as parameter set A. In addition to A, one can define another parameter set, B, which for the same set of parametric relations, solves the Jeans equation with the smallest residuals. 

{ We can explain the meaning of set B as follows. Suppose you were told that the equilibrium halos had Einasto-like density profiles, pseudo-phase space density profiles had power-law shape as a result of some physical principle, and anisotropy-density slope relation was approximately linear. You were then asked to obtain the parameter values characterizing these relations. The obvious way for you to proceed would be to solve the Jeans equation incorporating these relations, and obtain the best fitting set of parameters; that would be set B.}
%If you were then told that your parameter set differs from the one from simulations (set A), you would conclude that one of the relations you used is not universally true and should have been excluded. 

It is not a foregone conclusion that A and B are the same set. First, let us consider a situation---possibly hypothetical---where eq.~\eqref{ppsd} {is a property of Newtonian dynamical evolution, and characterizes all equilibrium halos.} The premise of our analysis is that in this case, the two sets of parameters will be the same, i.e. evolution will find the parameter set for which $\rho/\sigma^3$ is as close to a power law as it can be. On the other hand, {if eq.~\eqref{ppsd} does not have a physical origin, parameters sets A and B need not be the same, and the fact that eq.~\eqref{ppsd} appears to be satisfied in simulated halos is a coincidence.}

To assess the similarity of the two parameter sets, A and B, we evaluate the quality of solutions from a wide and finely sampled region of the model parameter space. Since all 3 relations, (i)-(iii), are of equal importance, one should treat them equally. This is not possible if one assumes exact forms for (i) and (ii), and then integrates the Jeans equation to get (iii). Fortunately, there is a way to place (i)-(iii), and their associated 4 or 6 model parameters, on the same footing. In section~\ref{2nd} we show that the second derivative of the Jeans equation, combined with an Einasto profile, can be converted to a cubic algebraic equation. We then calculate how well each set of parameters satisfies this algebraic equation.  In contrast to the integration of the Jeans equation, the algebraic equation does not single out $\rho/\sigma^3$.

 Since (i)-(iii) have been observed in simulations, we know that a rough agreement between sets A and B is guaranteed. Therefore, in order to provide support for eq.~\eqref{ppsd} having a physical meaning, we are looking for a better than a rough agreement. In that case, we also expect set B to form a well defined and isolated trough in the parameter space, and be stable against small changes in parametrization, like changing from 4 to 6 parameters.

{We note that our entire analysis is done with the assumption of spherical symmetry of halos, and is one of the limitations of our modelling. Real dark matter halos, even geometrically spherically symmetric equilibrium halos can have velocity structures that are not fully described by radial velocity anisotropy \citep{woj13}. However, most (if not all) analysis of equilibrium halos in N-body simulations is done {\em after} spherically averaging their properties, like density, radial and tangential velocity dispersions, and pseudo phase-space density. Because the results of these analyses form the starting point of our analysis, we are necessarily confined to the case of spherical symmetry.}

%% file: Sections/ACJE.tex
In this section we will work with the second derivative of the Jeans equation, and parametrized forms of the density profile, velocity anisotropy profile, and power law profile of pseudo phase-space density. 

We start with the anisotropic Jeans equation,
\begin{equation} \label{JE}
\frac{d}{dr}\left\{\frac{\rho(r)\sigma^2(r)}{3-2\beta(r)}\right\} + \frac{2\beta(r)}{3-2\beta(r)}\frac{\rho(r)\sigma^2(r)}{r} = -G\rho(r)\frac{M(r)}{r^2}
\end{equation}
where $M(r)$ is the mass enclosed within radius $r$, $\rho$ is the density at that radius, $\sigma$ is the total velocity dispersion\footnote{While some authors use just the radial velocity dispersion in eq.~\eqref{ppsd}, we use the total dispersion, $\sigma^2=\sigma_r^2+\sigma_\phi^2+\sigma_\theta^2.$}, and $\beta$ is the anisotropy, defined using the tangential, $\sigma_t^2=\frac{1}{2}(\sigma_\phi^2+\sigma_\theta^2)$, and radial velocity dispersions of a dark matter halo: $\beta(r)\equiv 1-\frac{\sigma_t^2}{\sigma_r^2}$. 
We define dimensionless variables $x \equiv r/r_0$ and $y \equiv \rho/\rho_0$, and reduce the number of functions in eq.~\eqref{JE} by invoking the power law nature of the pseudo phase-space density profile, $\rho/\sigma^3 = (\rho_0/\sigma_0^3)(r/r_0)^{-\alpha}$:
%##########################################################
%algebraic Jeans Eqn
%#########################################################
\begin{equation} \label{JEM}
 \frac{-x^2}{y} \left\{ \frac{d}{dr} \left[ \frac{y^{5/3}x^{2\alpha/3}}{3-2\beta(r)} \right] + \frac{2\beta(x)}{3-2\beta(x)} y^{\frac{5}{3}}x^{\frac{2\alpha}{3}-1}\right\} = BM(x)
\end{equation}
where $B = G/r_0v_o^2$. Using the assumption of eq.~\eqref{ppsd} leads to Jeans equation constrained. From now on, ACJ refers to anisotropic constrained Jeans equation. 

\subsection{Second derivative of the ACJ equation}\label{2nd}

We differentiate eq.~\eqref{JEM} with respect $x$, as was done in \cite{Tay01}, then following \cite{Wil04} and \cite{Bar07}, we differentiate it again with respect to $x$, arriving at
\begin{multline}\label{CJE}
    (2\alpha + \gamma -6)[\frac{2}{3}(\alpha-\gamma)+1](2\alpha-5\gamma)\\ =15\gamma''+3\gamma'(8\alpha-5\gamma+4\beta+12\beta\theta b_1 -5)\\
    -3\theta [b_1(4\alpha^2 +\gamma^2-8\alpha\gamma+8\alpha+7\gamma-15)]\\
    -3\theta^2[6b_1b_2(\alpha-\gamma+1)] \color{white}+alignmentedit\\
    -3\theta^3[b_3(54\beta+144\beta^2+24\beta^3)] \color{white}+aligned\\
    -3\theta'[6b_1(\alpha-\gamma+1)+9b_1b_2\delta]-3\theta''(3b_1)\color{white}xxxxxxxxxxxxxx
\end{multline}
%##########################################################
where $\gamma=-d \ln y/d \ln x$ is the double logarithmic density slope, $\theta=d \ln\beta/d \ln x$, $b_1=2\beta/(3 - 2\beta)$, $b_2=(3 + 2\beta)/(3 - 2\beta)$, $b_3=(3 - 2\beta)^{-3}$, and the primes indicate logarithmic derivatives. Using a range of initial conditions for collisionless N-body simulations, \cite{Han06a,Han06b} find that after equilibrium is achieved, all the halos end up having very similar shapes of the velocity anisotropy. All are well described by a linear relation\footnote{Note that the sign in front of $\eta_2$ is different from that in \cite{Han06a} and \citep{Han06b} because our definition of $\gamma$ has a minus sign, while theirs does not.} between $\beta$ and $\gamma$: 
\begin{equation}\label{betaH05}
\beta(r) = \eta_1 + \eta_2\gamma(r).
\end{equation}
This allows us to compute the logarithmic derivatives of $\theta$:
%##########################################################
%Theta derivatives
%#########################################################
\begin{subequations} \label{theta}
\begin{align}
\theta & = -\frac{\eta_2}{\beta}\gamma' \\
\theta' & = -\frac{\eta_2}{\beta} \left \{ \frac{\eta_2}{\beta}(\gamma')^2 + \gamma'' \right\}\\
\theta'' & = -\frac{\eta_2}{\beta} \left \{ \frac{2\,(\gamma')^2 \, \eta_2^2 \, \gamma}{\beta^3} + 2\gamma' \gamma'' + \frac{\gamma''}{\beta^2}\eta_2\gamma +\frac{\gamma'''}{\beta} \right\}
\end{align}
\end{subequations}

Eq.~\eqref{CJE} is a second order differential equation in $\gamma$ and $\theta$. It can be further expanded by using logarithmic derivatives in eq.~\eqref{theta}, thereby converting the eq.~\eqref{CJE} into a third order differential equation in one variable only, $\gamma$. Due to the complexity of eq.~\eqref{CJE} and \eqref{theta}, we are not showing the combined expression obtained after assembling all the parts together.  

\subsubsection{Single Einasto algebraic (SEA) equation}

The logarithmic density $\gamma$ can take many forms; a simple analytical expression that fits N-body density profiles well and is commonly used in the literature is the Einasto profile \citep{Ein65,Nav04}: 
\begin{equation}\label{einastoeq}
\gamma = Ax^p,
\end{equation}
where $A$ is a normalization constant. It has an interesting, and useful property that all its logarithmic derivatives have a linear dependence on $\gamma$:
%fix the alignment 
\begin{subequations} \label{Einasto}
\begin{align}
%\gamma &= Ax^p \\
\gamma' &= Apx^p = p\gamma \\
\gamma'' &= Ap^2x^p =  p^2\gamma\\
\gamma''' &= Ap^3x^p = p^3\gamma.
\end{align}
\end{subequations}
We take advantage of this property of the Einasto profile. In eq.~\eqref{CJE} and \eqref{theta} we eliminate all derivatives of $\gamma$ by replacing them with their counterparts in eq.~\eqref{Einasto}. This converts differential eq.~\eqref{CJE} into an algebraic equation of cubic order. From now on, we will refer to it as the single Einasto algebraic (SEA) equation that consists of a collection of eq.~\eqref{CJE} - \eqref{Einasto}. It depends on 4 parameters: the power law slope of the pseudo phase-space density $\alpha$, the two velocity anisotropy parameters, $\eta_1$, $\eta_2$ and the Einasto parameter $p$.

\subsubsection{Triple Einasto algebraic (TEA) equation}

Though Einasto profiles fit N-body dark matter halo density profiles quite well, they are still only fitting functions. DARKexp models have the advantage of being theoretically derived from fundamental statistical mechanical principles. They also fit the density profiles, and the energy distributions of N-body halos very well. It is not surprising that they resemble Einasto profiles for a certain range of DARKexp shape parameter $\phi_0$. DARKexp density profiles do not have an analytic expression, but since for the relevant range of $\phi_0$ they are not too different from the Einasto shape, DARKexp can be approximated by three Einasto segments, with three different slopes  $p_1$, $p_2$ and $p_3$, respectively. The profiles we use cover 3 decades in radius, from $x=10^{-2}$ to $x=10^1$, and $x=1$ corresponds to the radius where $\gamma=2$, i.e. has the isothermal slope. {Note that the shape of the profile at the very center, at $x<10^{-2}$ is irrelevant since it is excluded from the analysis.}
%##########################################################
%Density Profile Gamma
%#########################################################
\begin{align} \label{gamma}
    \gamma(x)= 
\begin{dcases}
    2\times 10^{1.3(p_1-p_2)}\, x^{p_1},&  \log x\leq -1.3\\
    2\,x^{p_2},                    & -1.3 \leq \log x\leq 0\\
    2\,x^{p_3},                    & 0\leq\log x.
\end{dcases}
\end{align}
%#########################################################
The continuity conditions are derived at $\log x = -1.3$ and $\log x = 0$ by forcing the density slopes $\gamma$ to be equal at the these two points, with the help of a constant multiplier in the logarithmic units. We also require that $p_2 \leq p_1$, in accordance with the shape of DARKexp density profiles \citep{Wil10b}. 

When eq.~\eqref{gamma} is substituted in to the eq.~\eqref{CJE}, the final equation becomes a function of 6 model parameters: the power law slope of the pseudo phase-space density $\alpha$, the two velocity anisotropy parameters, $\eta_1$, $\eta_2$, and three Einasto parameters $p_1$, $p_2$ and $p_3$, which collectively approximate DARKexp. We call it the triple Einasto algebraic (TEA) equation, which consists of eq.~\eqref{CJE} - \eqref{gamma}.

\subsubsection{Figure of merit for evaluating the solutions}\label{fom}

If the simple analytical assumptions---eq.~\eqref{ppsd},~\eqref{betaH05}, and eq.~\eqref{einastoeq} or eq.~\eqref{gamma}---are exactly correct, then the SEA and TEA equations will be exactly true at all radii, $x$. However, this is not the case for any set of 4 or 6 parameters. To estimate how well any given set of parameters solves the two equations over the 3 decades in radius, {we divide this entire range into 30 logarithmic intervals,} calculate the normalized absolute difference between the left hand side (LHS) and the right hand side (RHS) of the equations, and then average these values. {We chose log intervals because that is what is used by other researchers when fitting density and pseudo phase-space profiles to N-body simulated halos.}
The resulting value, which we call $\delta$, is defined as,
%##########################################################
%Delta Def
%#########################################################
\begin{equation} \label{delta}
\delta = \left\langle \frac{|LHS - RHS|}{\sqrt{(LHS^2 + RHS^2)/2}}\right\rangle_{{\log}x}
\end{equation}
%#########################################################

After scanning a large portion of the parameter space, we will identify regions that represent best solutions of the two equations (global minima of $\delta$) and compare the corresponding parameter values to those found in computer simulations. 

\subsection{Grid-based search for approximate solutions}\label{grid}

%##########################################################
%Figure : Heatmaps (Single Einasto)
%#########################################################

\begin{figure*}
    \centering
    \makebox[0pt][c]{
    \includegraphics[scale=0.5]{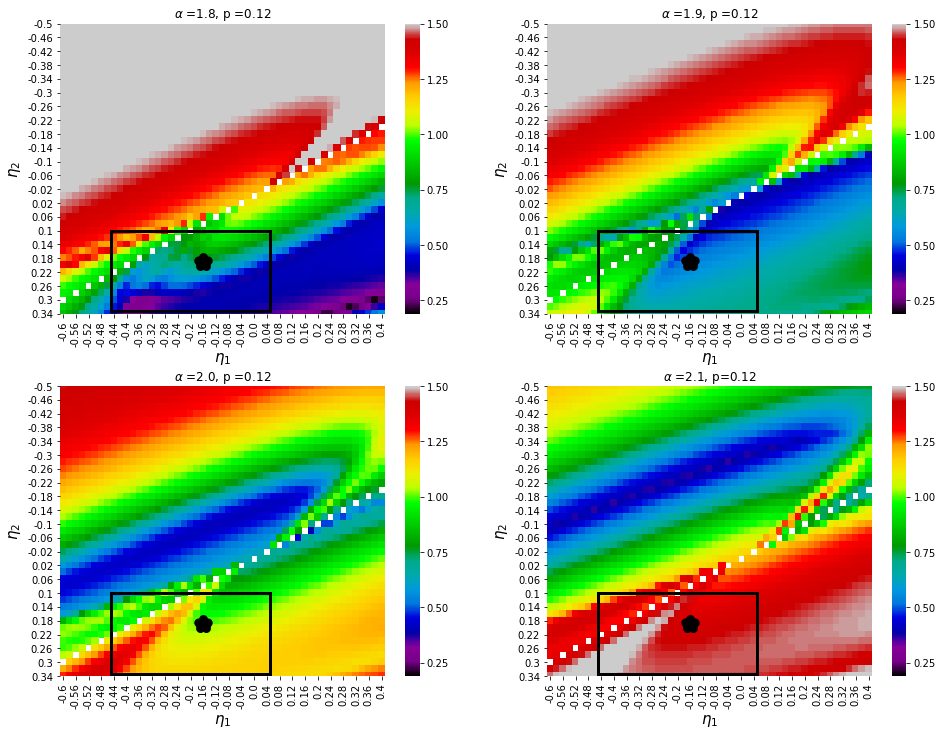}
    }
    \caption{\textit{Heat maps of $\delta$ values as a function of velocity anisotropy parameters, $\eta_1$ and $\eta_2$, for a single Einasto density profile, eq.~\eqref{einastoeq},  with the density slope value, $p = 0.12$, corresponding to the lowest $\delta$. The four panels show four values of $\alpha$ values, $1.8$, $1.9$, $2.0$ and $2.1$. The black rectangular regions outline $\eta_1$ and $\eta_2$ values obtained in \protect\cite{Han06a}, and are the limits of regions plotted in figure~\ref{fig:heatmaps}. The black dot marks the parameters presented in \protect\cite{Han06b}.}}
    \label{fig:heatmapsSE}
\end{figure*}

%##########################################################

\subsubsection{SEA equation}\label{solve1ACJE}

Here, we consider the single Einasto profile of eq.~\eqref{einastoeq}, resulting in a four dimensional parameter space. The parameter ranges covered are: $1.8\leq\alpha\leq2.2$, $-0.60 \leq \eta_1 \leq 0.40$, $-0.5 \leq \eta_2 \leq 0.36$, $0.02\leq p\leq 0.3$, with  step size of $0.02$ for all parameters. Our parameter range for $\alpha$ values is wider than what is found in the literature, which generally span $\sim 1.85-1.96$. The range of density profile slopes is also somewhat wider. The velocity anisotropy parameter ranges encompass those found in N-body simulations, as analyzed in \cite{Han06a}.

We start by calculating the figure of merit, $\delta$, eq.~\eqref{delta}, at every location in the four dimensional parameter space. Given our step size and parameter limits, we consider a total of around $5 \times 10^5$ parameter sets. The parameter sets with the lowest $\delta$ values correspond to the best solutions to the SEA equation. Figure of merit $\delta$ is always greater than zero, meaning that no solution is exact. As was pointed out in section~\ref{outline}, the advantage of using the second derivative of the Jeans equation is that, by design, it contains no information about which one of the three analytic assumptions---density, anisotropy, or $\rho/\sigma^3$ profile, or what combination of the three---is the reason for $\delta$ being non-zero.  In the rest of this subsection we study the interdependence of these parameters, examine the properties of the best solutions, and compare them to those of dark matter halos found in simulations.

\begin{figure*}
\centering 
\subfloat[]{%
  \includegraphics[width=0.9\columnwidth]{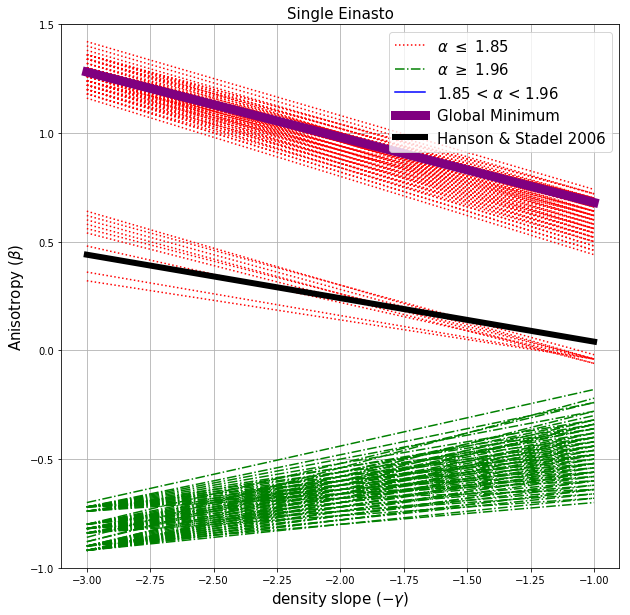}%
  \label{fig:fig_a}%
}\qquad
\subfloat[]{%
  \includegraphics[width=0.9\columnwidth]{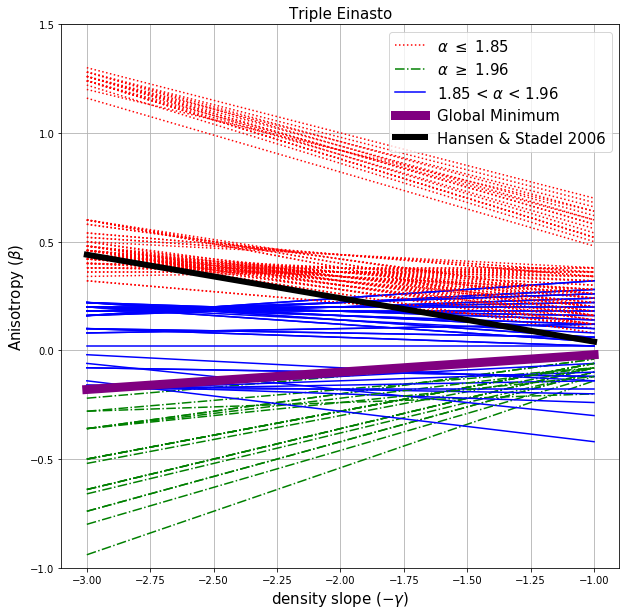}%
  \label{fig:fig_b}%
}
\caption{\textit{The anisotropy-density relation $\beta$ for the lowest 150 $\delta$ values found in the (a) 4 dimensional parameter space of the single Einasto profile, and (b) hexa-dimensional space of the triple Einasto profile. The black line in both panels assumes the anisotropy parameters from \protect\cite{Han06b}, of simulated dark matter halos. The thick purple lines are the global minima of our two searches. The profiles are colored by their value of $\alpha$, as indicated in the legend.}}
\end{figure*}

%##########################################################
%##########################################################

To display the solution space, we chose the density profile slope $p$ corresponding to the lowest global $\delta$ value ($p = 0.12$), and plot 4 cuts through the remaining 3D parameter space, corresponding to 4 values of $\alpha$: $1.8$, $1.9$, $2.0$ and $2.1$; see figure~\ref{fig:heatmapsSE}.  The color scale in these heat maps is the same for all 4 panels. The gray color indicates {very large} values of $\delta$, outside our range. The black box outlines the anisotropy parameters given in \cite{Han06a}, while the black dot marks the single set presented in \cite{Han06b}; both are based on simulated halos.

The pattern revealed in these approximate solutions is complicated. There is a symmetry axis outlined by white pixels, where the denominator of $\delta$ is zero for some values of $x$. It goes roughly diagonally through all 4 panels, and the color pattern is inverted across the axis. The trough of minima is to the right of this axis for $\alpha=1.8-1.9$, and shifts to the left for large values of $\alpha$.

%##########################################################
%Figure : Heatmaps (Triple Einasto)
%#########################################################

\begin{figure*}
    \centering
    \makebox[0pt][c]{
    \includegraphics[scale=0.5]{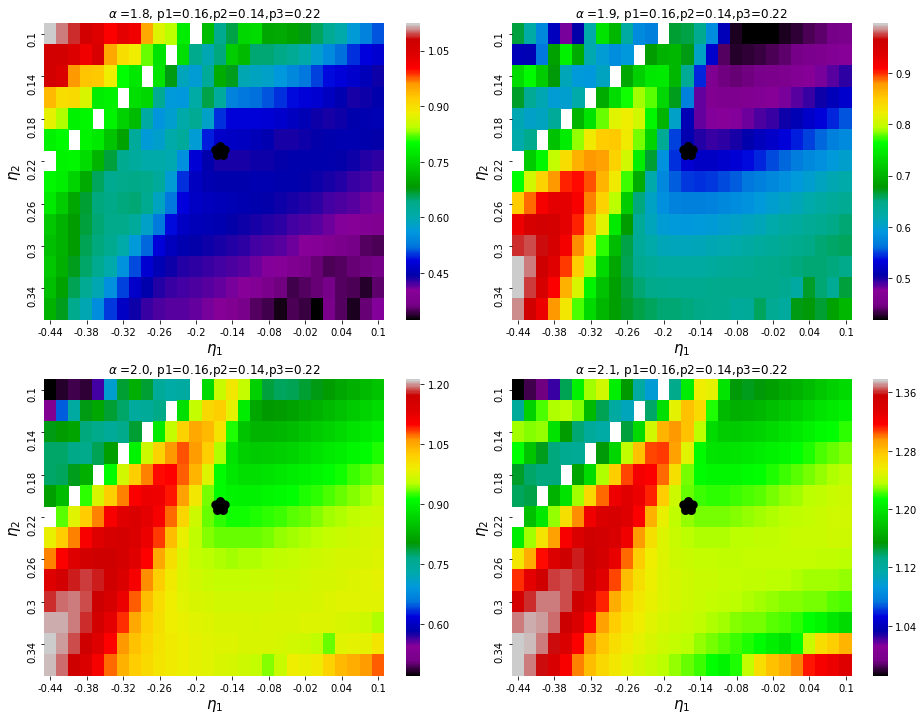}
    }
    \caption{\textit{Similar to figure~\ref{fig:heatmapsSE}. Heat maps of $\delta$ values as a function of velocity anisotropy parameters, $\eta_1$ and $\eta_2$, for a triple Einasto density profile, eq.~\eqref{gamma}, with the density slope values (indicated above each panel) corresponding to DARKexp with $\phi_0=4.5$. The density profiles are shown in figure~\ref{fig:densityslopeHansen}.  The parameter sets presented here are approximately those of set A. These values are not those of the lowest $\delta$ (set B), which are outside this range.  Note that the color ranges are different in the four panels. {(Some maps appear a little noisy at the level of the 3rd significant digit of $\delta$. Although $\delta$ is derived from an analytic equation, eq.~\eqref{CJE}, it has a complex dependence on anisotropy parameters, $\eta_1$ and $\eta_2$, and so need not necessarily predict very smooth behaviour.)}}}
    \label{fig:heatmaps}
\end{figure*}

%##########################################################

The global $\delta$ minimum is at $(\alpha,p,\eta_1,\eta_2)=(1.80,0.12,0.38,0.30)$ {i.e., set B for the SEA equation}, and does not correspond closely to the parameters of the simulated halos. Specifically, the anisotropies are too large, even exceeding $\beta=1$ (which is not allowed because of the definition of $\beta$), the Einasto density profile slope, $p$, is too small, and the slope of the pseudo phase-space density profile, $\alpha$ is a little shallower than the range seen in simulations. 

{To quantify the difference between set B parameters and those obtained in N-body simulations, set A, we use \cite{Lud11}, who find that the slope of the Einasto density profile slope, $p$, and the exponent of the pseudo phase-space density, $\alpha$, are linearly correlated; see their Figure 7. We estimate the uncertainty, $\sigma$, in these set A values from that figure: the dispersion around the linear relation is $\sigma_{p\alpha}\approx 0.015$. Here, $p\alpha$ is not a product, but is meant to represent that the two parameters are correlated. We also assume that the two anisotropy parameters are not correlated. Their uncertainty is taken to be half the range given in \cite{Han06a}, i.e., for the anisotropy parameter $\eta_1$, the uncertainty is $\sigma_{\eta_1}=0.27$, and for $\eta_2$, it is $\sigma_{\eta_2}=0.12$. 

For case SEA we estimate the significance of the difference between set A and set B parameters by vectorizing
\begin{equation*}
    \sigma_{SEA} = \left(\frac{|{p\alpha_A - p\alpha_B}|}{\sigma_{p\alpha}},\frac{|{\eta_{1_A} - \eta_{1_B}}|}{\sigma_{\eta_1}},\frac{|{\eta_{2_A} - \eta_{2_B}}|}{\sigma_{\eta_2}}\right),
\end{equation*}
\label{eqsigma}
where $|p\alpha_A-p\alpha_B|$ is the distance in the plane of $p$ vs. $\alpha$ of our set B parameters and the closest set A parameters in Figure 7 of \cite{Lud11}. We find that $\sigma_{SEA} = 7.4$, hence the two sets are significantly different from each other.}

The best 150 solutions in the space of anisotropy-density relation are displayed in figure~\ref{fig:fig_a}. The global minimum is the thick purple line, and is very different from the anisotropy-density relation found in simulations. The thin lines (150 solutions) are color coded by their $\alpha$ value. The range found in simulations, $1.85\leq \alpha\leq 1.96$ (blue) does not occur in this set. 

The fact that the parameter set corresponding to the global $\delta$ minimum (set B) does not coincide with the one observed in simulations (set A) is an indication that $\rho/\sigma^3$ is not as close to a power law as it could have been, and hence does not have a physical significance.  { This conclusion is insensitive to the exact shape of the anisotropy-density relation, and remains valid even if the relation is only roughly linear. This is because the global $\delta$ minimum is outside of the range obtained from simulations (black box in figure~\ref{fig:heatmapsSE}).}

% {\sout{As we argued in section~\ref{outline}, if $\rho/\sigma^3$ being a power law had a physical origin, the lowest $\delta$ solution would coincide closely with the parameter set found in simulations, and there would have been a trough of low $\delta$ values around the latter parameter set, as one would expect for any well defined minimum in a parameter space.}}

\subsubsection{TEA equation}\label{solve3ACJE}

In this section we extend the range of density profiles we consider, by including profiles that resemble DARKexp, which can be approximated by three joined Einasto-like segments, eq.~\eqref{gamma}. 

The corresponding hexa-dimensional parameter space we search spans the same range for $\alpha$, $\eta_1$ and $\eta_2$ as in section~\ref{solve1ACJE}. The values of the slope of the two Einasto segments of the density profile at smaller radii, $p_1$ and $p_2$, cover the same range as $p$, but the segment that applies to larger radii spans $0.16\leq p_3\leq0.24$, because the density profile is expected to be steeper there. We impose an additional constraint that $p_2\leq p_1$, as indicated by the shape of DARKexp density profiles. The step size in each of the 6 dimensions is $0.02$, so the total number of parameter sets we consider is nearly $2.5\times 10^7$.

Figure $\ref{fig:heatmaps}$ is similar to figure~\ref{fig:heatmapsSE}, but only shows $\eta_1$ and $\eta_2$ ranges that are indicated by black rectangles in the latter figure. This is the velocity anisotropy range found in simulations \citep{Han06a}. The 4 panels correspond to the same four $\alpha$ values as in that figure, while the 3 values of the density profile slopes of triple Einasto are {those that resemble the density profile of DARKexp with $\phi_0=4.5$.}

Just like the $\delta$ maps in figure~\ref{fig:heatmapsSE}, the ones in figure~\ref{fig:heatmaps} also show complicated patterns. (Note that the color scale is different in the four panels.)  The global $\delta$ minimum, i.e. our set B, is at $(\alpha,p_1,p_2,p_3,\eta_1,\eta_2)=(1.94,0.2,0.18,0.16,0.06,-0.08)$. Its parameter values are not the same as those found in simulations. The $\alpha$ value is a little too large, the density profile is only marginally well approximated by either Einasto or DARKexp, and the anisotropy $\beta$ is mildly tangentially anisotropic at large radii %(purple line in figure~\ref{fig:Anisotropy}), 
in contrast to the radially anisotropic velocity profiles of simulated halos. %(black line). 
{We quantify the difference between set A and set B parameters using the same standard deviation analysis described in section~\ref{solve1ACJE}. 
Since published papers on N-body simulations present only single, not triple Einasto fits, we use the same result from \cite{Lud11} for case TEA.  This is acceptable because set B values for $(p_1,~p_2,~p_3)$ are very close to each other, $(0.20,~0.18,~0.16)$, so we can use the average, $p=0.18$. With this, set A and set B differ at $3.3\sigma$ level.}

Figure~\ref{fig:fig_b} shows the anisotropy-density relation for the 150 best solutions. The global minimum is shown as a thick purple line. The average $\beta-\gamma$ relation from simulations \citep{Han06b} is plotted for reference as the thick black line. Most, if not all of the best 150 solutions have properties different from those of simulated equilibrium halos. All green lines, and the thick purple line have the anisotropy-density slope opposite to that seen in simulations, and $\alpha$ values that are larger than those seen in simulations; $\alpha\geq 1.96$. The solutions represented by red lines have $\alpha$ values that are too small, $\alpha\leq 1.85$. Furthermore, many of these have $\beta>1$ in the relevant range of density slopes. The solutions that have $\alpha$ in the observed range, $1.85\leq \alpha\leq 1.96$ (blue lines) have approximately isotropic velocity distributions at all radii, and are thus only marginally consistent with simulations. 

\begin{figure*}
\centering 
\subfloat[]{%
  \includegraphics[width=0.9\columnwidth]{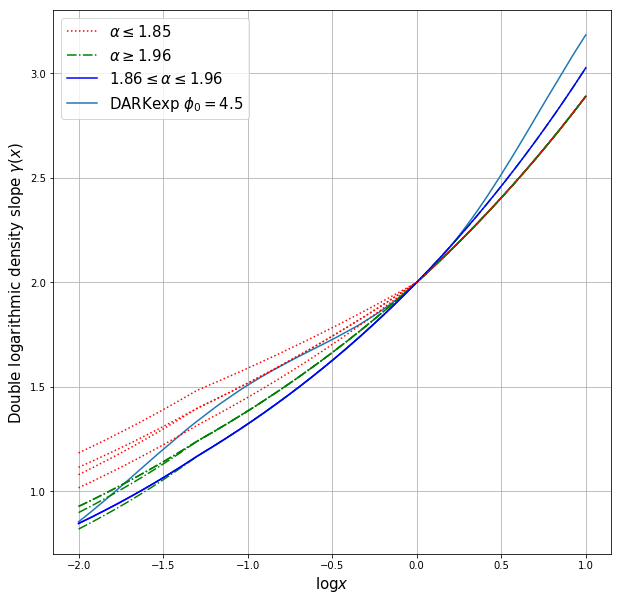}%
  \label{fig:densityslope}%
}\qquad
\subfloat[]{%
  \includegraphics[width=0.9\columnwidth]{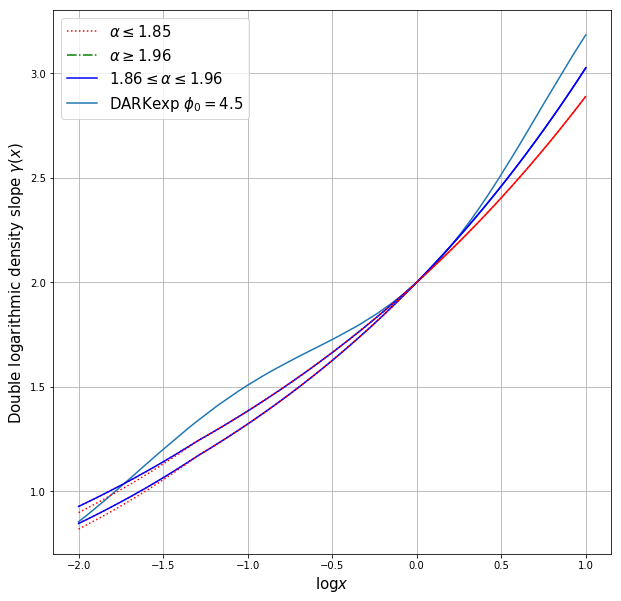}%
  \label{fig:densityslopeHansen}%
}
\caption{\textit{Double logarithmic density slope $\gamma$ as a function of logarithmic scaled radius, $\log(x)$. In both panels, the thin lines are the 150 best solutions from section~\ref{solve3ACJE}, color coded by $\alpha$. DARKexp $\phi_0=4.5$ density profile is plotted as light blue. (a) Solutions, regardless of anisotropy parameters; (b) Solutions with anisotropy parameters in the \protect\cite{Han06a} range. There is considerable overlap of the curves, so in all cases a single profile actually represents many profiles.}}
\end{figure*}

Figure~\ref{fig:densityslope} shows the density profile slopes vs. log radius, for the 150 best solutions, color coded by $\alpha$ as in figure~\ref{fig:fig_b}, and figure~\ref{fig:densityslopeHansen} plots the subset of these 150 solutions that are within \cite{Han06a} anisotropy range. Though we allowed a rather broad range of slopes for $p_1$, $p_2$ and $p_3$, the best solutions are all clustered around a narrow range of values, resulting in considerable overlap of profiles in the figure. These values are such that $p_1\approx p_2\approx 0.16-0.18$, and $p_3 \approx 0.18-0.20$. In other words, even though the density profile had the option of deviating from an Einasto form, the best solutions still have $p_1\approx p_2\approx p_3$. The closest DARKexp has $\phi_0=4.5$, and is represented as a light blue line. So the density profiles corresponding to best solutions are similar to, but not the same as those found in N-body simulations.

To sum up, the parameter set found in simulations is (i) somewhat similar to, but (ii) not very close to the best solution to the anisotropic constrained Jeans equation. The first statement is a posteriori conclusion, which was already shown to be the case by \cite{Tay01}. The second statement suggests that the power law nature of $\rho/\sigma^3$ found in simulations is a coincidence. { If eq.~\eqref{ppsd} had physical significance, nature would have found a different set of parameters---namely, our set B---to satisfy eq.~\eqref{ppsd}, while keeping density profiles Einasto or DARKexp-like, and anisotropy-density slope relation approximately linear.}

A further conclusion drawn from these results argues against eq.~\eqref{ppsd} having a physical significance. { If it were, one would expect the parameters of the single Einasto (section~\ref{solve1ACJE}) and triple Einasto (section~\ref{solve3ACJE}) best solutions to be similar. This is not the case.} While the density profiles in both cases are Einasto-like (for the triple Einasto $p_1\approx p_2\approx p_3$), the slopes are different, and the power law exponents $\alpha$, and the velocity anisotropies are very different. Thus, a relatively small change in the parametrization of the problem significantly changes the parameters of the global minimum. {The unstable solution suggests that eq.~\eqref{ppsd} is not a universal feature of dynamical evolution.}

%% file: Sections/Conclusion.tex
It has been known for almost 20 years that the pseudo phase-space density profiles of equilibrium dark matter halos are well approximated by a power law in over $\sim 3$ decades radius. The main goal of this paper is to determine whether this scale-free behavior could have a physical origin, or is simply a curious coincidence. While we do not address the possible physical meaning of $\rho/\sigma^3$ being a power law, we assume that if one exists, equilibrium halos will obey the eq.~\eqref{ppsd} relation.

We work with the final equilibrium halos, instead of analyzing their dynamical evolution. We use the Jeans equation of hydrostatic equilibrium, and the three simple parametric relations that describe the radial behavior of three profiles: density, velocity anisotropy profile, and $\rho/\sigma^3$, which are  parametrized by a total of 4 or 6 model parameters, depending on whether the density is described by a single or triple Einasto profiles. We then search the parameter space, over a wide and finely sampled range, for the best solution to the Jeans equation {(set B)}. Because all three relations are seen in simulations, we already know that parameters describing N-body halos {(set A)} will correspond to reasonably good solutions of the Jeans equation. However, if $\rho/\sigma^3\propto r^{-\alpha}$ has a physical origin, we expect N-body halos to correspond to the best solutions {(i.e., set A and B to be the same within uncertainties)}, and expect these to form a well defined, isolated, and stable global minimum trough. 

What we actually find is quite different. The structure of the solution space is complicated, with no indication of an isolated trough. {The parameter set found in N-body simulations (set A) and the one that corresponds to a global minimum we find (set B) are different at $\sim 3-7\sigma$ significance. Since numerical work like ours can explore only a limited range of parameters, and will necessarily leave large portions of parameter space unexplored, we need to consider the possibility that the true global minimum (set B) is outside of our range. That means that set B is even further away from set A, which only strengthens our conclusion that the two sets are not the same.}

Furthermore, we find that if the parametrization of the density profile is changed somewhat, from a single to triple Einasto, the parameters of the best solution, and especially those of velocity anisotropy change significantly indicating that the global minimum in the solution space is not stable. 

Because the parameters describing the density, velocity dispersion and $\rho/\sigma^3$ radial profiles observed in simulations are not the ones that would result in the $\rho/\sigma^3$ being closest to a power law, we conclude that the approximate power law nature of the pseudo phase-space density seen in N-body simulations and semi-analytical collapses does not have a physically meaningful origin, and so does not shed light on the effective equation of state of self-gravitating dark matter halos. 

%{ Finally, we note that the second derivative of the Jeans equation, first introduced by \cite{Wil04,Aus05}, can have other useful applications, beyond the purpose of this paper.
%Defining  $f=$ LHS$-$RHS, where the left and right hand sides refer to the Jeans equation itself, one can Taylor expand around any radius $x_0$,
%\begin{equation}
%f(x)=f(x_0)
%+(x-x_0)\,\frac{\partial f}{\partial x}\Big\vert_{x_0}
%+\frac{1}{2}(x-x_0)^2\,\frac{\partial^2 f}{\partial x^2}\Big\vert_{x_0}+\,\,...
%\end{equation}
%In equilibrium, $f$ and all its derivatives are zero, because the Jeans equation is satisfied. The second derivative of the Jeans equation, $\partial^2 f/\partial x^2$, which is related to the 3rd derivative of the gravitational potential, and, unlike the Jeans equation and its first derivative does not depend on the halo mass normalization, could serve as a metric for halo stability. Its usefulness in this capacity can be evaluated from the results of high resolution N-body simulations, by calculating it at different radii, and at different epochs during evolution. If $\partial^2 f/\partial x^2$ proves to be a useful metric for stability, it can be employed in the search for possible dynamical attractors during the later stages of the collapse and relaxation of dark matter halos.